\documentclass[twocolumn, pra, showpacs,superscriptaddress]{revtex4-2}

\usepackage{graphicx}     % 插入图片
\usepackage{dcolumn}      % 表格列对齐
\usepackage{bm}           % 粗体数学符号
\usepackage{hyperref}     % 交叉引用超链接
\usepackage{xcolor}       % 颜色处理（可选）
\usepackage{amsmath}   
\usepackage{epstopdf}
\usepackage[caption=false]{subfig}
\usepackage{amsmath, amssymb, amsfonts}
\usepackage{placeins}
\newcommand{\ket}[1]{|#1\rangle}
\newcommand{\bra}[1]{\langle#1|}

\begin{document}
	
	% --- 标题与作者信息 ---
	\title{Antibunching Enhancement via Non-Markovianity in a Hybrid Optical-Microwave Cross-Cavity}
	
	\author{Shiying Gu}
	\affiliation{Institute of Theoretical Physics and State Key Laboratory of Quantum
		Optics Technologies and Devices, Shanxi University, Taiyuan 030006, China}
	\author{Jun-Qi Li}
	\affiliation{Institute of Theoretical Physics and State Key Laboratory of Quantum
		Optics Technologies and Devices, Shanxi University, Taiyuan 030006, China}
	\author{Liping Guo}
	\email{guolp@sxu.edu.cn}
	\affiliation{Institute of Theoretical Physics and State Key Laboratory of Quantum
		Optics Technologies and Devices, Shanxi University, Taiyuan 030006, China}
	\date{\today}
	
	% --- 摘要 ---
	\begin{abstract}
		% 请在此处插入您的英文摘要内容
	Unconventional photon blockade (UPB) provides an attractive route for generating antibunched light through quantum interference without requiring strong nonlinear coupling. In conventional Markovian systems, irreversible dissipation progressively destroys the phase coherence, even limiting the achievable single-photon purity. Here we propose a scheme for imposing non-Markovian effects on a microwave cavity where a Er$^{3+}$:Y$_2$SiO$_5$ crystal loaded into a hybrid optical-microwave system. Based on the time-convolutionless non-Markovian framework, we derive the time-dependent non-Markovian decay rate and the renormalized microwave drive in the weak-coupling regime. We show that the second-order correlation function is significantly suppressed, even to the order of $10^{-7}$, since the backflow of environmental information enhances photon blocking. In addition, this structured reservoir engineering not only provides an additional degree of freedom, but also further optimizes the blockade via optical detuning. Importantly, the delayed second-order correlation exhibits a broadened antibunching profile, thereby relaxing the timing requirements for single-photon synchronization. Our results demonstrate that environmental non-Markovianity can be transformed from a source of decoherence into a controllable quantum resource, opening new opportunities for high-purity single-photon generation.
	\end{abstract}
	
	\maketitle
	
	% --- 正文部分 ---
	\section{Introduction}
	Generating high-quality single-photon sources is a fundamental objective in quantum information processing \cite{aharonovich2016, senellart2017}.
One well-established approach is photon blockade (PB) \cite{PhysRevA.46.R6801, PhysRevLett.79.1467}, where the quantum effect of a single photon inhibits the absorption or transmission of subsequent photons, yielding nonclassical light with sub-Poissonian statistics. The PB has been successfully implemented across various platforms, including cavity quantum electrodynamics (QED) \cite{bin2018, grujic2012, hamsen2017, hou2019, leboite2016, peyronel2012, radulaski2017, rosenblum2011, shen2015, shen2015a, wu2024, zhang2023a, zou2020}, quantum dots \cite{faraon2010, foster2019, muller2015, shen2020, snijders2018, zhang2014}, optomechanical systems \cite{liao2013, rabl2011, sarma2018, wang2015}, and circuit QED \cite{fink2017, flayac2015, hoffman2011, liu2014, miranowicz2013, vaneph2018}. It is generally classified into two categories: the conventional photon blockade (CPB) requires strong spectral anharmonicity and nonlinearity exceeding dissipation rates \cite{birnbaum2005, leoifmmodeacutenelsenfiski1994}; while unconventional photon blockade (UPB) achieves strong antibunching via destructive interference pathway in weakly nonlinear systems \cite{bamba2011, liew2010}. Beyond conventional single-photon generation, the principles of the PB have inspired investigations into broader quantum optical phenomena, including nonreciprocal light transmission \cite{huang2018, li2019a, liu2023, shen2020a, wang2019a, wu2025, xie2022}, non-Hermitian dynamics associated with parity-time ($\mathcal{PT}$) symmetry \cite{li2015, liu2024, wang2019} and exceptional points \cite{ben-asher2023, huang2022, yuan2023, zhou2018} and multi-photon blockade \cite{zhu2017, zou2020a}.

In general, any realistic quantum system inevitably interacts with its surrounding environments \cite{devega2017}. While such open-system is conventionally treated under the Markovian approximation, this scheme does not have perfect second-order correlation, and even fails when the environmental correlation time is not negligibly  \cite{breuer2016, li2019}. In contrast,  non-Markovian dynamics is characterized by a backflow of information from the environment, which manifests as memory effects \cite{xue2012, zhang2017}. These effects fundamentally originate from diverse physical sources, for example, structured environmental spectral densities, nonlocal correlations between environmental degrees of freedom, and correlations in the initial system-environment state \cite{breuer2016}. Thus, non-Markovianity has also emerged as a powerful tool for advancing practical applications like quantum-state engineering and coherent control \cite{PhysRevA.97.062104}. In quantum optics, actively tailoring the system-environment coupling to exploit these memory effects offers a promising avenue to protect and manipulate quantum states \cite{verstraete2009}.

A prototypical configuration to realize the multipath interference is the three-level atom system. Previous studies have demonstrated UPB under both far-detuned or resonant conditions \cite{dong2024, gao2023,  li2019, shen2023, sun2026, wang2020, you2024, zhang2023, zou2022}. However, in standard Markovian frameworks, inherent system dissipations degrade the phase coherence necessary for destructive interference. To suppress this decoherence, existing methods typically introduce auxiliary resources, such as degenerate parametric amplifier (DPA) \cite{PhysRevA.90.063824} or additional classical drives \cite{wu2024, haider2023}. Alternatively, utilizing the memory effects of a structured environment has been demonstrated in driven dissipative coupled-cavity systems without relying on external active components in Yi X. X Group \cite{PhysRevA.109.043714}. Inspired by this, we propose one model of a hybrid optical-microwave crossed cavity, where the microwave resonator functions as a structured reservoir. In our theoretical treatment, the non-Markovian backflow is modulated by the second-order correlation function, which serves as the memory kernel in a Volterra integro-differential equation to manipulate the system's decoherence evolution \cite{nakajima1958, zwanzig1960}. Our study highlights enhancing photon antibunching well beyond the Markovian steady-state limit. Furthermore, the transient non-Markovian blockade exhibits an extended temporal correlation feature, which relaxes timing constraints for single-photon synchronization and demonstrates the potential of structured reservoirs for scalable quantum photonic applications.

The remainder of this paper is organized as follows: In Sec. \ref{sec:model}, we introduce a model of the hybrid optical-microwave crossed cavity and the system Hamiltonian. In Sec. \ref{sec:non}, we establish the non-Markovian framework to derive the memory kernel, the decoherence function, and TCL master equation, where reservoir feedback is represented by a time-dependent decay rate $\gamma_{{\rm NM}}(t)$ and microwave drive $\Omega_{m}(t)$. In Sec. \ref{TRANSIENT}, we investigate the transient photon blockade and reveal the density matrix interference pathways responsible for the emergence of non-Markovian UPB. In Sec. \ref{PARAMETER}, we present the numerical results and discuss the relevant physical mechanisms. Finally, in Sec. \ref{sec:con}, we summarize our main conclusions.
	\section{Model}
	\label{sec:model}
\begin{figure}[htbp] 
		\centering
		\subfloat[]{
			\includegraphics[width=0.45\textwidth]{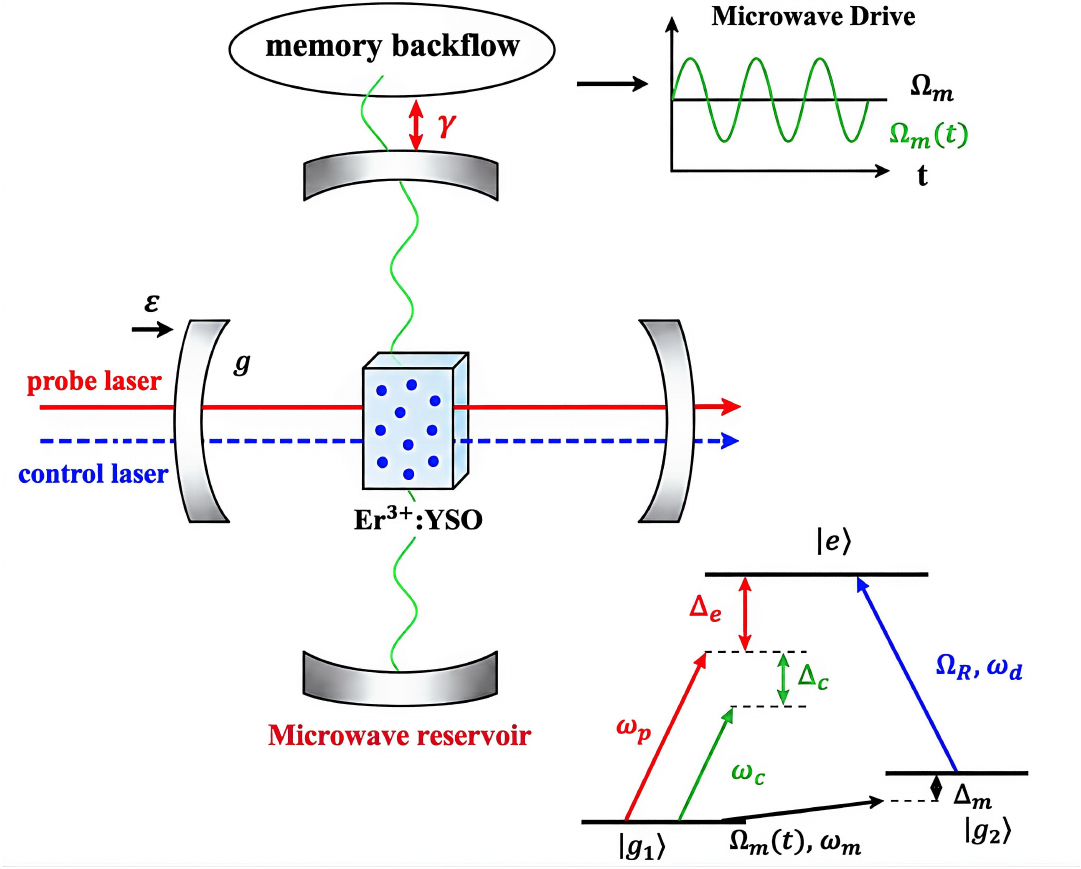
			}
			\label{fig1a}
		}
		\hfill 
		\subfloat[]{
			\includegraphics[width=0.4\textwidth]{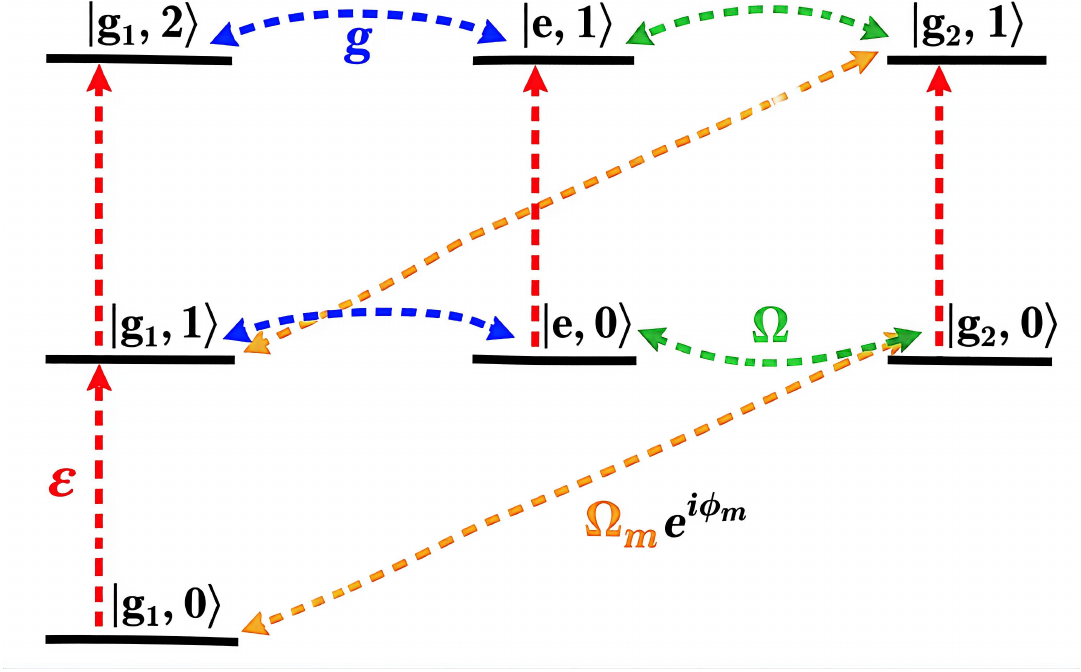}
			\label{fig1b}
		}
		\caption{Schematic of the crossed optical-microwave cavity and the level configuration. (a) An Er$^{3+}$:Y$2$SiO$5$ crystal is positioned at the intersection of an optical Fabry-Pérot cavity and a microwave resonator. The lower right inset illustrates the $\Delta$-type three-level atomic configuration, where $|g_1\rangle$ and $|g_2\rangle$ are the hyperfine ground states, and $|e\rangle$ is the excited state. The top right inset is that the constant bare amplitude $\Omega_m$ evolves into a time-dependent effective field $\Omega_m(t)$ due to the non-Markovian environmental feedback. (b) Energy-level and transition pathways for the destructive quantum interference}.
		\label{fig1}
	\end{figure}

Specifically, we consider an effective closed-loop $\Delta$-type three-level system implemented in an Er$^{3+}$: Y$_2$SiO$_5$ crystal, which is positioned at the intersection of an optical Fabry-Pérot cavity and a metallic microwave resonator (Fig. \ref{fig1}) \cite{Adwaith:19}. This hybrid cross-cavity is designed for convenient operation, simultaneously realizing Markovian dissipation in the optical regime and non-Markovian backflow in the microwave regime. The two lower states, $\vert{}g_{1}\rangle$ and $\vert{}g_{2}\rangle$, are selected from the Zeeman or hyperfine sublevels of the $^4I_{15/2}$ ground-state manifold of the Er$^{3+}$ ions, while the excited state $\vert{}e\rangle$ belongs to the $^4I_{13/2}$ manifold. Specifically, along the optical path, the transition $\vert{}g_{1}\rangle \leftrightarrow \vert{}e\rangle$ is coupled to a cavity mode with frequency $\omega_{c}$ and coupling strength $g$, while being simultaneously driven by a weak external probe field with frequency $\omega_{p}$ and amplitude $\varepsilon$. The adjacent optical transition $\vert{}g_{2}\rangle \leftrightarrow \vert{}e\rangle$ is driven by a classical control field with Rabi frequency $\Omega_{R}$ and angular frequency $\omega_{d}$. Crucially, compared to conventional atomic-vapor implementations, this solid-state Er$^{3+}$:Y$_2$SiO$_5$ platform avoids Doppler broadening and provides long-lived spin sublevels. These properties construct an ideal environment for preserving the fragile multi-path quantum interference required for UPB.

The ground-state magnetic-dipole transition $\ket{g_{1}}\leftrightarrow\ket{g_{2}}$ is addressed by a microwave cavity with frequency $\omega_{m}$, Rabi-frequency $\Omega_{m}$ and phase $\phi_{m}$.  The Hamiltonian of the system is given by
	\begin{equation}
		\begin{aligned}
			\hat{H}_{0} = & \omega_{c}\hat{a}^{\dagger}\hat{a} + \omega_{e}\ket{e}\bra{e}+ \omega_{2}\ket{g_{2}}\bra{g_{2}} + g(\hat{a}\hat{\sigma}_{e1} + \hat{a}^{\dagger}\hat{\sigma}_{1e}) \\
			& + \Omega_{R}(\hat{\sigma}_{e2}e^{-i\omega_{d}t} + \hat{\sigma}_{2e}e^{i\omega_{d}t})+  \varepsilon(\hat{a}^{\dagger}e^{-i\omega_{p}t} + \hat{a}e^{i\omega_{p}t})\\
			& + \Omega_{m}(\hat{\sigma}_{21}e^{-i\omega_{m}t} + \hat{\sigma}_{12}e^{i\omega_{m}t}),
		\end{aligned}
	\end{equation}
	 where level $\ket{g_{1}}$ is set as the zero potential energy point. The first three terms represent the free Hamiltonians of the cavity and the atom, and here $\hat{a}$ ($\hat{a}^{\dagger}$) is the bosonic annihilation (creation) operator. The atomic projection operators are defined as $\hat{\sigma}_{ij}=\ket{i}\bra{j}$ with $i,j=1,2,e$.	
	In a rotating frame defined by $\hat{U}(t)=\exp[i(\omega_{p}\hat{a}^{\dagger}\hat{a} + \omega_{p}\hat{\sigma}_{ee} + \omega_{m}\hat{\sigma}_{22})t]$ with the resonance condition $\omega_{p}=\omega_{d}+\omega_{m}$, the system Hamiltonian becomes:
	\begin{equation}
		\begin{aligned}
			\hat{H}^{'}_{0} = & \Delta_{c}\hat{a}^{\dagger}\hat{a} + \Delta_{e}\ket{e}\bra{e} + \Delta_{m}\ket{g_{2}}\bra{g_{2}} \\
			& + g(\hat{a}\ket{e}\bra{g_{1}} + \text{h.c.})+ \Omega_{R}(\ket{e}\bra{g_{2}} + \text{h.c.}) \\
			& + \Omega_{m}(\ket{g_{2}}\bra{g_{1}}e^{i\phi_{m}} + \text{h.c.}) + \varepsilon(\hat{a}^{\dagger} + \text{h.c.}),
		\end{aligned}
		\label{2}
	\end{equation}
		Where the detunings are $\Delta_{c}=\omega_{c}-\omega_{p}$, $\Delta_{e}=\omega_{e}-\omega_{p}$, and $\Delta_{m}=\omega_{2}-\omega_{m}$. Enabled by the crossed-cavity architecture, the system interacts with two distinct reservoirs. The transitions undergo standard Markovian decay induced by a flat  spectrum. Next, the microwave transition $\ket{g_{1}} \leftrightarrow \ket{g_{2}}$ couples to a structured, finite-bandwidth reservoir \cite{lambropoulos2000}, triggering pronounced non-Markovian information backflow \cite{devega2017}. 
		\section{NON-MARKOVIAN OPEN-SYSTEM DYNAMICS}
		\label{sec:non}
		\subsection*{A.Memory Kernel}
	To characterize the structured non-Markovian environment, we model the microwave reservoir with $\hat{H}_{B}=\sum_{k}\omega_{k}\hat{b}_{k}^{\dagger}\hat{b}_{k}$, where $\hat{b}_{k}$ and $\hat{b}_{k}^{\dagger}$ denote the annihilation and creation operators of the $k$-th harmonic oscillator mode, respectively. Here the interaction between the atom and the reservoir is described by the interaction Hamiltonian $\hat{H}_{I}=(|g_{2}\rangle\langle g_{1}| + \text{H.c.})\otimes\hat{B}$,  where this coupling operator $\hat{B}$ can be expanded as a linear superposition of the reservoir modes:
		\begin{equation}
	\hat{B}=\sum_{k}(g_{k}\hat{b}_{k}+g_{k}^{*}\hat{b}_{k}^{\dagger})
		\end{equation}
	  where $g_{k}$ represents the corresponding coupling strength between the system and the $k$-th harmonic mode. Transforming $\hat{B}$ into the rotating Heisenberg picture with respect to the microwave driving frequency $\omega_m$ yields:
	  \begin{equation}
	\hat{B}(\tau) = \sum_k \left[ g_k \hat{b}_k e^{-i(\omega_k - \omega_m)\tau} + g_k^* \hat{b}_k^\dagger e^{i(\omega_k - \omega_m)\tau} \right].\quad
	\end{equation}	
		 To evaluate the environmental correlation, we assume that the microwave environment is in thermal equilibrium at temperature $T$,  the density matrix $\rho_{B}=e^{-\beta\hat{H}_{B}}/Z_{B}$ (with $\beta=1/(k_{B}T)$ and $Z_{B}$ the partition function), the symmetric correlation function of the environment is defined as:
	\begin{equation}
	f(\tau)=\frac{1}{2}\left(\langle\hat{B}(\tau)\hat{B}(0)\rangle+\langle\hat{B}(0)\hat{B}(\tau)\rangle\right),
\end{equation}
	where $\langle\cdot\rangle=\text{Tr}_{B}\{\rho_{B}\cdot\}$ denotes the thermal equilibrium average. The key contributing term is the mean thermal photon number $n_{k}=(e^{\beta\hbar\omega_{k}}-1)^{-1}$. Then by introducing the environmental spectral density $J(\omega)=\sum_{k}|g_{k}|^{2}\delta(\omega-\omega_{k})$, the correlation function $f(\tau)$ can be expressed as a continuous frequency integral. To align with the rotating frame and facilitate analytical evaluations using complex variable theory, we extend the integration domain to the entire real axis and transform the integration variable to the relative frequency $\Delta = \omega - \omega_m$ (see Appendix \ref{A} for detailed derivations). This yields the correlation function:
	\begin{equation}
	f(\tau)=\frac{1}{2}\int_{-\infty}^{+\infty}d\Delta J(\Delta)\coth\left(\frac{\beta\hbar\omega_{mc}}{2}\right)e^{i\Delta\tau}.
 \end{equation}
	Here, the Lorentzian spectral density \cite{tanimura2006,groblacher2015}:
		\begin{equation}
J(\Delta)=\frac{2\gamma^{2}\Delta_{mc}\lambda\Delta}{(\Delta^{2}-\Delta_{mc}^{2})^{2}+\lambda^{2}\Delta^{2}}
\end{equation}
where $\Delta_{mc} = \omega_{mc} - \omega_m$ is the detuning between the bare microwave cavity frequency $\omega_{mc}$ and the driving frequency $\omega_m$. It is crucial to note that while the coherent oscillations and spectral profile of the correlation function are entirely determined by the rotating-frame frequency $\Delta$, and the $\coth$ factor encodes the absolute physical frequency $\omega_{mc}$. Furthermore, $\lambda$ denotes the spectral width, whose inverse is proportional to the environmental memory time. $\gamma$ represents the effective atom-reservoir coupling strength. Evaluating $f(\tau)$ via the residue theorem, the poles of the integrand consist of both the complex poles of $J(\Delta)$ and the imaginary Matsubara poles originating from the thermal factor. In the high-temperature limit ($\beta\hbar\omega_{mc}\ll1$), the contribution from the Matsubara terms decays exponentially as $e^{-2\pi\tau/\beta\hbar}$. Since this decay is significantly faster than the memory decay of the reservoir, it can be safely neglected within the relevant timescale of the non-Markovian dynamics. Ultimately, we obtain the fully analytical non-Markovian memory kernel:
		\begin{equation}
	f(\tau)=K e^{-\lambda\tau/2}\begin{cases}
		{\displaystyle \cos\left(\Omega\tau\right)+\frac{\lambda}{2\Omega}\sin\left(\Omega\tau\right),} & \Delta_{mc}>\lambda/2,\\[1em]
		{\displaystyle 1+\frac{\lambda}{2}\tau,} & \Delta_{mc}=\lambda/2,\\[1em]
		{\displaystyle \cosh\left(\Omega\tau\right)+\frac{\lambda}{2\Omega}\sinh\left(\Omega\tau\right),} & \Delta_{mc}<\lambda/2,
	\end{cases}
	\label{8}
\end{equation}
	where $K={2\pi\gamma^{2}}/{\beta\hbar\omega_{mc}}$ characterizes the effective system-environment coupling strength, and $\Omega=\sqrt{\Delta_{mc}^{2}-\lambda^{2}/4}$ signifies the effective oscillation frequency.
	Eq. \eqref{8} reveals that the dynamical evolution of the memory kernel $f(\tau)$ is governed by the competition between $\Delta_{mc}$ and $\lambda$. In the underdamped regime ($\Delta_{mc}>\lambda/2$), $f(\tau)$ exhibits sinusoidal oscillations modulated by a global exponential decay envelope $e^{-\lambda\tau/2}$. By appropriately tuning the system parameters, these oscillations can cross zero, leading to periodic negative excursions of the environmental correlation function. Consequently, the energy and coherence previously dissipated into the reservoir flow back into the atomic system, giving rise to non-Markovian  backflow. Conversely, in the critically damped or overdamped regimes ($\Delta_{mc}\le\lambda/2$), these oscillatory features degenerate into monotonic linear or hyperbolic decays, marking an irreversible information loss that approaches the Markovian limit. Therefore, we choose the reservoir parameters within the underdamped regime. It is emphasizing that our chosen temperature is $T = 20\text{ K}$ and the equivalent effective thermal occupation $n_{th}\simeq60$. We will show that with such thermal noise, the UPB effect is protected and revived by the non-Markovian memory effects .
		\subsection*{B. Decoherence Function}
	The core of non-Markovian open systems is the memory-induced  feedback that strongly modulates  decoherence. To  track this temporal evolution, we introduce the Green's function $G(t)$, acting as the decoherence function to characterize the probability amplitude of the system retaining its initial quantum coherence in the non-Markovian process \cite{garraway1997}. In contrast to the monotonic exponential decay inherent to the Markovian approximation, the evolution of $G(t)$ encompasses not only dissipation but also information backflow and coherence revival driven by the continuous exchange of microwave photons between the atom and the reservoir.  The decoherence process is  governed by the Volterra integro-differential equation \cite{breuer2002theory}:
	\begin{equation}
	\dot{G}(t)=-\int_{0}^{t}d\tau f(\tau)G(t-\tau)
\end{equation}
	with the initial condition $G(0)=1$. The convolution term on the right-hand side intuitively encapsulates the quintessential feature: the rate of  $\dot{G}(t)$ at the present time $t$ depends not exclusively on the current system state, but on the accumulated history of past states $G(t-\tau)$, weighted by the memory kernel $f(\tau)$. To circumvent the analytical difficulties by this non-local temporal convolution, we apply the Laplace transform to obtain the frequency-domain algebraic solution:
		\begin{equation}
	\tilde{G}(s)=\frac{1}{s+\tilde{f}(s)}
	\label{10}
\end{equation}
where $\tilde{f}(s)=K(s+\lambda)/[(s+\lambda/2)^{2}+\Omega^{2}]$. Although this frequency-domain solution is  concise (See Appendix \ref{B} for details), returning to the time domain is imperative to comprehensively capture the transient dynamics. Therefore, we recast Eq. \eqref{10} into an ordinary differential equation (ODE). Specifically, we introduce an auxiliary variable $y(t)=-\dot{G}(t)$ to equivalently represent the convolution term. Via the inverse Laplace transform, the original dynamics are  mapped onto a second-order linear ODE:
	\begin{equation}
\ddot{y}+\lambda\dot{y}+(\Delta_{mc}^{2}+K)y-K\lambda G=0.
\label{11}
\end{equation}
On one hand, numerical integration of Eq. \eqref{11} can be utilized to directly yield the transient decoherence function $G(t)$, this implicit approach obscures critical time-dependent features like decay rates and frequency shifts. On the other hand, to pursue an analytical expression, we analyze the characteristic polynomial from the denominator of Eq. \eqref{10}: $P(s)=s(s^{2}+\lambda s+\Delta_{mc}^{2})+K(s+\lambda)$. Its three complex roots, $s_{j}$, dictate the intrinsic system dynamics. Based on partial fraction expansion, $\tilde{G}(s)$ can be universally expressed as a superposition of simple poles, $\tilde{G}(s)=\sum_{j}C_{j}/(s-s_{j})$, $C_{j}$ are the respective residues, regardless of the coupling strength $K$. To bypass the complexity of cubic equations, we assume the weak-coupling condition $K\ll\Delta_{mc}^{2}$ and treat $K(s+\lambda)$ as a perturbation to the bare  $P_{0}(s)=s(s^{2}+\lambda s+\Delta_{mc}^{2})$ (see Appendix \ref{B} for calculation details). This yields the reconstructed analytical decoherence function:
\begin{equation}
	\begin{aligned}
		G(t) &= e^{-\frac{K\lambda}{\Delta_{mc}^{2}}t} + K \Bigg[ \frac{2\Omega-i\lambda}{\Omega(2\Omega+i\lambda)^{2}} e^{(-\frac{\lambda}{2}+i\Omega)t} \\
		&\quad + \frac{2\Omega+i\lambda}{\Omega(2\Omega-i\lambda)^{2}} e^{(-\frac{\lambda}{2}-i\Omega)t} \Bigg] + O(K^{2}).
	\end{aligned}
	\label{12}
\end{equation}
Here, the first term captures the long-time Markovian exponential decay with a purely dissipative relaxation rate $\Gamma_{M}=K\lambda/\Delta_{mc}^{2}$. The content in the bracketed of the second term encode the non-Markovian corrections from the structured reservoir, driving complex oscillations at frequency $\Omega$ under a memory decay envelope $e^{-\lambda t/2}$. Benefiting from the finite-bandwidth memory effects, the system experiences a backflow of dissipated energy and information from the environment, which manifests as a localized recovery of  coherence, providing a dynamic degree of freedom to lock in the destructive interference for UPB.
\subsection*{C. Unified TCL Master Equation and Drive Renormalization}
To comprehensively describe the transient dynamics of the atom within the hybrid environment, we incorporate both the non-Markovian effects and the  Markovian dissipation into a unified master equation framework. In Eq. \eqref{2} the structured microwave reservoir renormalizes the bare amplitude $\Omega_{m}$ into a time-dependent effective amplitude $\Omega_{m}(t)$. Since the optical and microwave dissipative channels are spatially independent and address orthogonal energy-level transitions, their respective contributions to system evolution satisfy the principle of linear superposition. Then at the same time $\hat{H}^{'}_{0}$ transforms into $\hat{H}^{'}_{0}(t)$. Consequently, the general form of the time-dependent master equation governing the reduced density matrix $\hat{\rho}(t)$ can be formulated as:
\begin{equation}
\frac{d\hat{\rho}}{dt}=-i[\hat{H}^{'}_{0}(t),\hat{\rho}]+\mathcal{L}_{opt}[\hat{\rho}]+\mathcal{L}_{NM}(t)[\hat{\rho}].
\label{13}
\end{equation}
 The optical transitions are governed by the standard time-independent Lindblad superoperator:
\begin{equation}
\mathcal{L}_{opt}[\hat{\rho}]=\kappa\mathcal{D}[\hat{a}]\hat{\rho}+\gamma_{1}\mathcal{D}[|g_{1}\rangle\langle e|]\hat{\rho}+\gamma_{2}\mathcal{D}[|g_{2}\rangle\langle e|]\hat{\rho}
\end{equation}
where $\kappa$ is the optical cavity decay rate, and $\gamma_{1,2}$ denote the spontaneous emission rates for the corresponding atomic transition branches, and the total spontaneous emission rate is $\gamma \equiv \gamma_1+ \gamma_2$. The critical task now translates to deriving the exact form of the time-dependent non-Markovian superoperator $\mathcal{L}_{NM}(t)$, and determining how the environmental memory dynamically renormalizes the microwave drive within the Hamiltonian $\hat{H}^{'}_{0}(t)$. To achieve this, we apply the time-convolutionless (TCL) master equation approach \cite{shibata1977,chaturvedi1979} to the  subspace ($|g_{1}\rangle\leftrightarrow|g_{2}\rangle$). This technique maps the non-Markovian evolution of the entangled system-environment onto an equivalent time-local differential equation, which allows us to distill the intricate environmental feedback into a time-dependent decay rate $\gamma_{NM}(t)$ and a renormalized drive $\Omega_{m}(t)$.
Drawing upon the formalism of dynamical maps, the evolution of the microwave subsystem can be written as:
\begin{equation}
\rho_{sub}(t)=G(t)\rho_{sub}(0)+I(t)
\end{equation}
Here, $I(t)=\int_{0}^{t}G(t-\tau)\Omega_{m}d\tau$ represents the accumulated inhomogeneous contribution originating from the microwave drive. Taking the time derivative and substituting the initial state relation $\rho_{sub}(0)=[\rho_{sub}(t)-I(t)]/G(t)$ to eliminate the dependence on $\rho_{sub}(0)$, we arrive at the subsystem master equation:
\begin{equation}
\dot{\rho}_{sub}(t)=\frac{\dot{G}(t)}{G(t)}\rho_{sub}(t)+\left[\dot{I}(t)-\frac{\dot{G}(t)}{G(t)}I(t)\right]
\label{16}
\end{equation}
Consequently, the time-dependent non-Markovian decay rate \cite{breuer2016} is naturally defined as
\begin{equation}
 \gamma_{NM}(t)=-\dot{G}(t)/G(t).
 \label{17}
 \end{equation}
 By substituting the first-order perturbative  solution of $G(t)$ into this definition, we obtain the explicit expression for the non-Markovian dissipation rate:
\begin{equation}
\gamma_{\mathrm{NM}}(t)=\frac{K\lambda}{\Delta_{mc}^{2}}+\frac{Ke^{-\frac{\lambda}{2}t}}{\Omega}\sin(\Omega t-2\phi)+O(K^{2})
\label{18}
\end{equation}
where $\phi=\operatorname{Arg}\left(4\Omega^{2}-\lambda^{2}+4i\lambda\Omega\right)/2$. The corresponding non-Markovian superoperator is thus given by
\begin{equation} \mathcal{L}_{NM}(t)[\hat{\rho}]=\gamma_{NM}(t)(\mathcal{D}[|g_{1}\rangle\langle g_{2}|]\hat{\rho}+\mathcal{D}[|g_{2}\rangle\langle g_{1}|]\hat{\rho}).
\end{equation}
To characterize the photon statistics during the transient evolution, we numerically solve this master equation to obtain the full density matrix $\rho_{{\rm N}}(t)$. The transient equal-time second-order correlation function is :
\begin{equation}
g^{(2)}(0,t)=\frac{Tr[\hat{n}(\hat{n}-1)\rho(t)]}{Tr[\hat{n}\rho(t)]^{2}},
\end{equation}
where $\hat{n}=\hat{a}^{\dagger}\hat{a}$ is the cavity photon number operator. The value of $g^{(2)}(0,t) < 1$ represents the photon antibunching ,which is corresponding to sub-Poissonian photon statistics.
 Additionally, the second term of Eq. \eqref{16} accounts for the reservoir-induced  renormalization of the microwave drive. Given that $I(t)=\int_{0}^{t}G(t-\tau)\Omega_{m}d\tau=\Omega_{m}\int_{0}^{t}G(\tau)d\tau$ and $\dot{I}(t)=\Omega_{m}G(t)$, renormalized microwave drive can be evaluated
\begin{equation}
	\Omega_{m}(t)=\Omega_{m}\left[G(t)+\gamma_{\text{NM}}(t)\int_{0}^{t}G(\tau)d\tau\right].
\label{19}
\end{equation}	
\begin{figure}[htbp]  
	\centering         
	\includegraphics[width=0.5\textwidth]{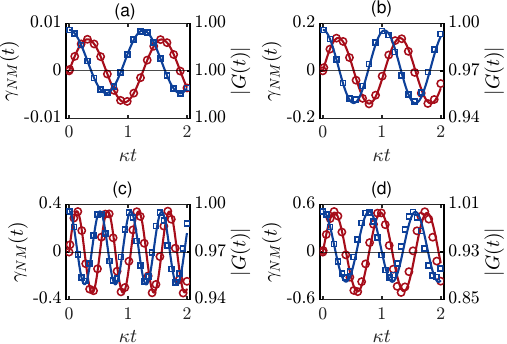}
	\caption{Comparison of the analytical and numerical solutions for the non-Markovian dissipation rate $\gamma_{\text{NM}}(t)$ (left axis, red) and the decoherence amplitude $|G(t)|$ (right axis, blue). Solid lines represent the first-order perturbative analytical solutions, while open circles and squares denote the numerical results obtained via the ODE method. (a) $\gamma=0.01$, $\Delta_{mc}=5$; (b) $\gamma=0.05$, $\Delta_{mc}=6$; (c) $\gamma=0.1$, $\Delta_{mc}=12$; and (d) $\gamma=0.1$, $\Delta_{mc}=8$.}
	\label{fig2}    
\end{figure}
So far, we have obtained the Green's function $G(t)$ and the non-Markovian dissipation rate $\gamma_{NM}(t)$ through numerical evaluation and analytical first-order perturbation expansion in $K$. Next, we investigate the parameter boundaries to determine where this analytical approximation remains valid. As shown as in Fig. \ref{fig2}(a)-\ref{fig2}(c), the perturbative and numerical results align perfectly. Fig. \ref{fig2}(b) and \ref{fig2}(c) verify the validity boundary of the analytical approximation at $K/\Delta_{mc}^{2}\approx0.0267$. Beyond this regime, as shown in Fig. \ref{fig2}(d) with $K/\Delta_{mc}^{2}\approx0.06$, the analytical approximation is no longer valid. Here the numerical curve exhibits a cumulative phase shift relative to the analytical prediction. This discrepancy stems from the environmental dressing effect: the first-order perturbation locks the evolution of the system at the bare frequency $\Omega$, thereby omitting the $K$-dominated higher-order corrections $\delta\Omega$. This omission leads to a phase drift $\Delta\phi(t)=\delta\Omega\cdot t$. These demonstrate that strict adherence to the weak coupling boundary $K\ll\Delta_{mc}^{2}$ is imperative to the validity of the analytical framework. In the subsequent analysis, we fix the condition for $\gamma=0.01$ and $\Delta_{mc}=5$.
\begin{figure}[htbp] 
	\centering
	\subfloat[]{
		\includegraphics[width=0.4\textwidth]{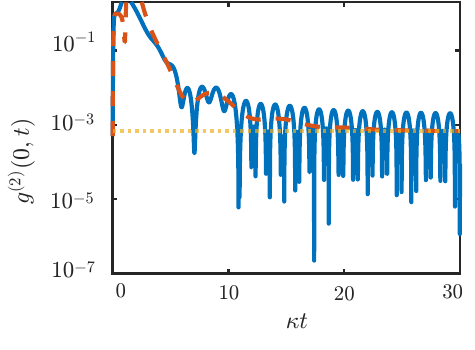}
		\label{fig3a}
	}
	\hfill
	\subfloat[]{
		\includegraphics[width=0.4\textwidth]{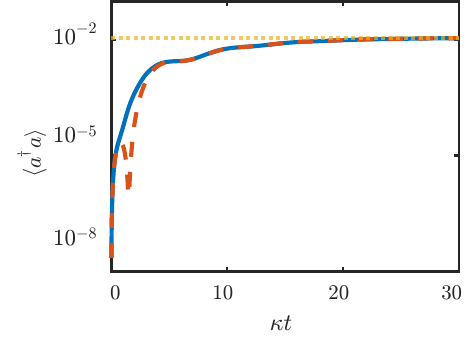}
		\label{fig3b}
	}
	\caption{Comparison results of the second-order coherence $\log_{10}[g^{(2)}(0,t)]$ (a) and the photon numbers $\langle a^\dagger a \rangle$ (b) given by the Markovian (red dashed line) and the non-Markovian evolution (blue solid line) respectively.  The yellow dotted line is the Markovian steady-state limit. The system parameters are set as $g=1.0\kappa$, $\Omega_R=0.5\kappa$, $\epsilon=\gamma=\lambda=0.01\kappa$, $\Delta_c=\Delta_e=0$, $\Delta_m=0.051\kappa$, $\Omega_m=0.035\kappa$, $\Delta_{mc}=5.0\kappa$ and $\phi_m=0$.}
	\label{fig3}
\end{figure} 
\section{TRANSIENT PHOTON BLOCKADE AND LIOUVILLE-SPACE INTERFERENCE}
 \label{TRANSIENT} 	
In this section, we demonstrate the emergence of transient UPB through exact numerical simulations, and unveil its origin via a weak-probe Liouville-space expansion. Fig. \ref{fig3a} shows the temporal evolution of the second-order coherence $g^{(2)}(0,t)$. Traditional Markovian systems eventually reach a static blockade. The non-Markovian dynamics exhibits a transient blockade feature, and this allows deeper destructive interference at specific instants, even driving $g^{(2)}(0,t)$ down to $2.2069\times10^{-7}$ (solid blue line) at $\kappa t$=17.4 in the first illustration of Fig. \ref{fig3a}. The evolution of the mean photon number $\left\langle a^{\dagger}a\right\rangle $ in Fig. \ref{fig3b} elucidates the energy-backflow mechanism.  Due to the initial Rabi exchange, the Markovian system exhibits a vacuum population. However, the non-Markovian curve traverses this stage smoothly. Notably, although the intrinsic evolution of the non-Markovian system exhibits strong oscillations, it eventually reaches the same stable brightness as the Markovian, indicating that the photon blockade governed by this mechanism holds practical value for signal output.

 \begin{figure}[htbp] 
 	\centering
 	\subfloat[]{
 		\includegraphics[width=0.4\textwidth]{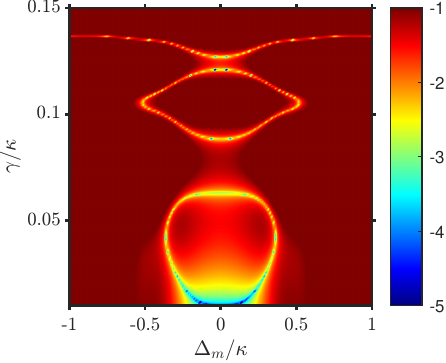}
 		\label{fig5a}
 	}
 	\hfill 
 	
 	\subfloat[]{
 		\includegraphics[width=0.4\textwidth]{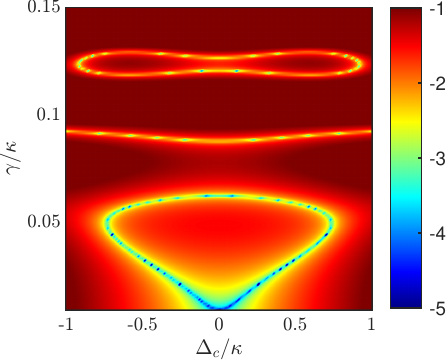}
 		\label{fig5b}
 	}
 	\caption{Transient photon blockade patterns and detuning dependence of the non-Markovian UPB at $t=19.5/\kappa$. The logarithmic second-order coherence $\log_{10}[g^{(2)}(0,t)]$ is shown as a function of (a) $\Delta_m/\kappa$ and $\gamma/\kappa$. Here $\Delta_c/\kappa=0$. (b) $\Delta_c/\kappa$ and $\gamma/\kappa$, here $\Delta_m/\kappa=0.051$. The other parameters chosen are the same as Fig. 3.}
 	\label{fig5}
 \end{figure}
 \begin{figure*}[htbp] 
 	\centering
 	\subfloat[]{
 		\includegraphics[width=0.35\textwidth, height=4.5cm, keepaspectratio=false]{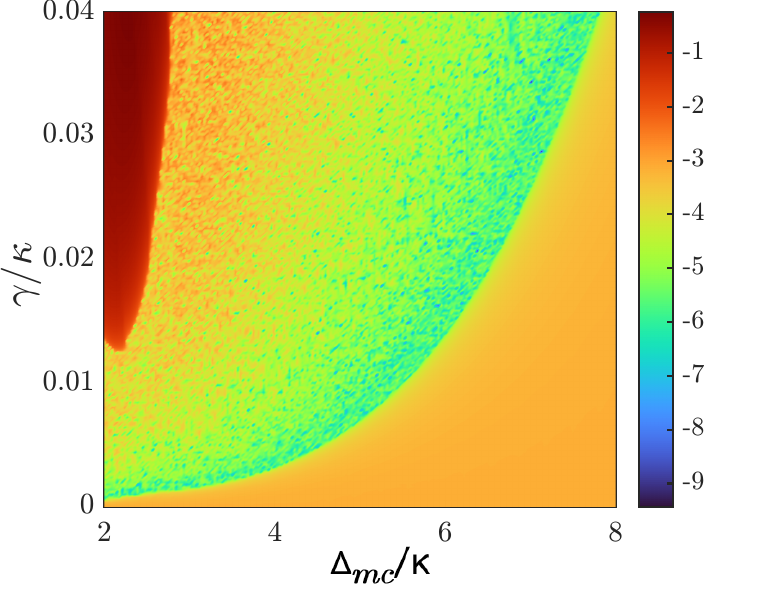}
 		\label{fig6a}
 	}\hfill 
 	\subfloat[]{
 		\includegraphics[width=0.3\textwidth, height=4.5cm, keepaspectratio=false]{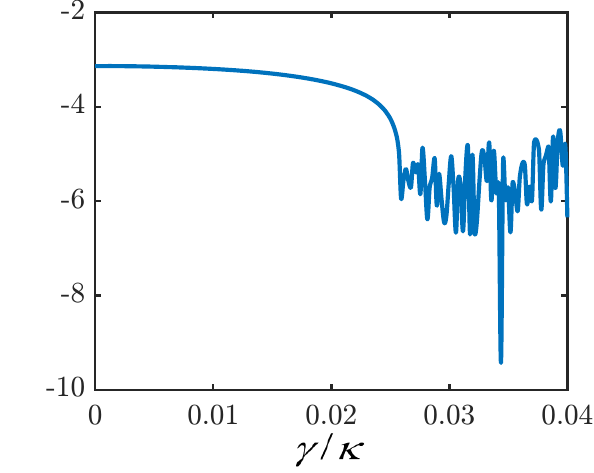}
 		\label{fig6b}
 	}
 	\subfloat[]{
 		\includegraphics[width=0.3\textwidth, height=4.5cm, keepaspectratio=false]{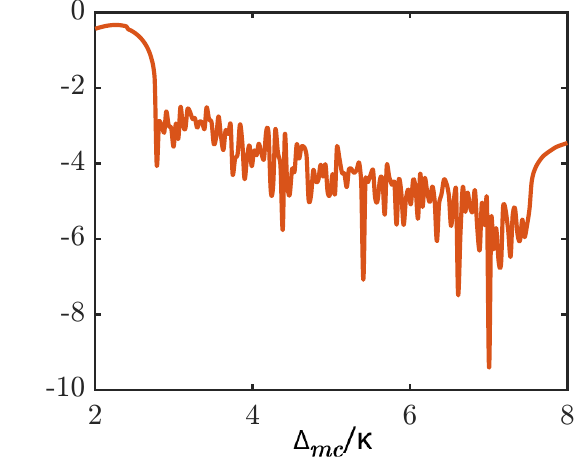}
 		\label{fig6c}
 	}\hfill 
 	\caption{Contour plot of transient minimum $\min_t \log_{10}[g^{(2)}(0, t)]$ as a function of $\Delta_{mc}/\kappa$ and $\gamma/\kappa$ (a), and panels (b) and (c) display the 1D cross-sections along the $\Delta_{mc} = 7.0\kappa$ and $\gamma= 0.034\kappa$ axes, respectively. Every data point in panel (a) is independently evaluated by continuously scanning $0 \le \kappa t \le 40$ to capture the optimal interference time $t_{opt}$. The other parameters chosen are the same as those in Fig. 3.}
 	\label{fig6}
 \end{figure*}
 \begin{figure}[htbp]  
 	\centering         
 	\includegraphics[width=0.45\textwidth]{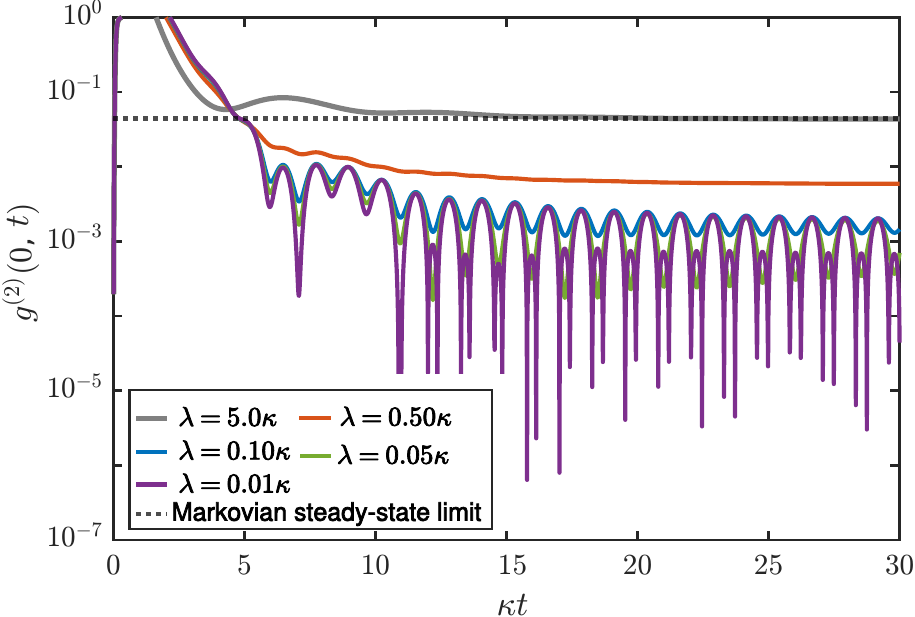}
 	\caption{Temporal evolution of $g^{(2)}(0, t)$ with various reservoir spectral widths $\lambda$, illustrating the transition from non-Markovian to Markovian dynamics. The colored solid lines plot the results for different values of $\lambda$, ranging from the non-Markovian regime to the Markovian limit ($\lambda/\kappa = 0.01,0.05,0.10,0.5, 5.0$). The black horizontal dotted line indicates the Markovian steady-state asymptote.}
 	\label{fig7}    
 \end{figure}
 \begin{figure}[htbp] 
 	\centering
 	\subfloat[]{
 		\includegraphics[width=0.83\columnwidth,
 		keepaspectratio]{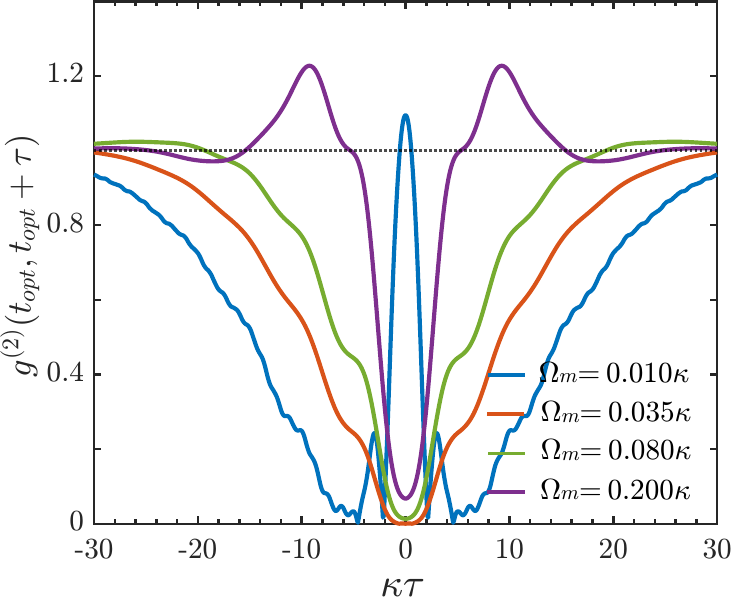}
 		\label{fig8a}
 	}
 	\hfill
 	\subfloat[]{
 		\includegraphics[width=0.83\columnwidth,
 		keepaspectratio]{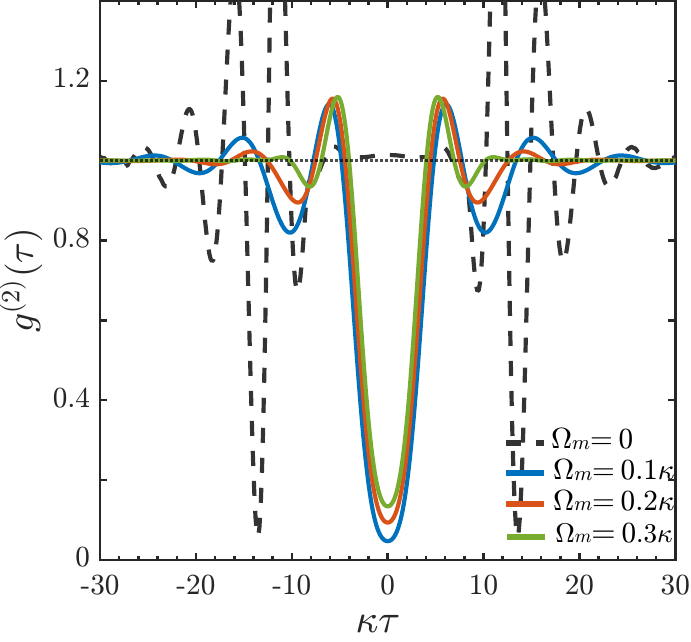}
 		\label{fig8b}
 	}
 	\caption{Delayed second-order correlation functions: (a) The transient non-Markovian correlation $g^{(2)}(t_{opt}, t_{opt}+\tau)$ evaluated at the optimal time $t_{opt}$. The colored solid lines plot the results for $\Omega_{m} = 0.010\kappa$, $t_{opt}=29.5\kappa$ (blue).  $\Omega_{m}=0.035\kappa$, $t_{opt}=28.3\kappa$ (red).  $\Omega_{m}=0.080\kappa$, $t_{opt}=10.0\kappa$ (green), and $\Omega_{m}=0.200\kappa$, $t_{opt}=10.0\kappa$ (purple). (b) The steady-state Markovian correlation $g^{(2)}(\tau)$. The lines represent $\Omega_{m} = 0$ (black dashed), $0.1\kappa$ (blue solid), $0.2\kappa$ (red solid), and $0.3\kappa$ (green solid). The black horizontal dotted lines in both panels indicate $g^{(2)}=1$. Other fixed system parameters are identical to those in Fig.3.}
 	\label{fig8}
 \end{figure}
 
Unlike the previously UPB method \cite{bamba2011}, in the present non-Markovian master equation the contributions of $\sigma_{12}\rho\sigma_{21}+\sigma_{21}\rho\sigma_{12}$ can not be ignored, since these terms continuously redistribute populations and coherences within the two ground-states. Here it is indispensable to our non-Markovian approach. To capture these mixed-state processes, we instead perform a weak-probe expansion directly in Liouville space (see Appendix \ref{D}), in which the density matrix is expanded to fourth order while retaining the complete dissipative dynamics. The expansion confirms that the transient antibunching originates from multi-order destructive interference in the mixed-state two-photon moment. To verify this interpretation, we numerically remove the recycling terms from the non-Markovian master equation, as shown in Appendix Fig. \ref{fig10}. This confirms that the observed UPB here is a dissipation-induced mixed-state interference driven by non-Markovian quantum jumps.
\section{PARAMETER SPACE OPTIMIZATION OF NON-MARKOVIAN UPB}
\label{PARAMETER}
With the interference mechanism established in the previous section, we now turn to the practical regulation and optimization of the non-Markovian UPB. In order to show the influence of the control detuning parameter, we plot the two-dimensional color scale map of $g^{(2)}(0,t)$ as a function of the microwave detuning $\Delta_{m}$ as well as the optical detuning $\Delta_{c}$ in Fig. \ref{fig5}. The color bar on the right-hand side represents the magnitude of $g^{(2)}(0,t)$. A distinctive feature is that both figures exhibit even symmetry about the resonance points $\Delta_{m}$ and $\Delta_{c} =0$. More importantly, single-photon blockade emerges at and near the resonance points in both figures,indicating that the non-Markovian feedback modifies the interference condition. Meanwhile, as $\gamma$ increases, the enhanced memory backflow is accompanied by stronger dissipation, progressively narrowing the detuning ranges supporting strong antibunching shrinks into a closed interference manifold. In the upper branch of the manifold in  Fig. \ref{fig5a}, as the coupling increases further, the closed manifold tears open into infinitely extending filaments. The underlying mechanism is that the Green's function approaches zero in this regime, triggering a negative divergent singularity ($\gamma_{\text{NM}} \to -\infty$), which is similar to the physical mechanism in the middle branch of Fig. \ref{fig5b}. In contrast, the cavity-detuning map in Fig. \ref{fig5b} reveals a cross-domain compensation mechanism, where the optical cavity detuning $\Delta_c$ provides an additional degree of freedom to recover the optimal condition shifted by the microwave reservoir. Interestingly, strong coupling here does not cause infinite divergence. A large cavity detuning blocks external photon injection, depleting its population.The closed boundary therefore marks the accessible parameter region within which cross-domain tuning can effectively restore the interference condition.
	
To evaluate the regulatory of structured reservoir on UPB, Fig. \ref{fig6a} presents the minimum of  $\log_{10}[g^{(2)}(0, t)]$ in the $0 \le \kappa t \le 40$, spanned by $\Delta_{mc}$ and $\gamma$. Here two distinct nonlinear boundaries emerge where Markovian behavior dominates in the inner and outer zones, while the middle zone exhibits a non-Markovian behavior, and the photon blockade is enhanced at the boundary directly bordering the outer region. Notably, antibunching region is constituted by a sparse grid of discrete points. This is not a numerical artifact, as $\Delta_{mc}$ and $\gamma$ are finely tuned, the optimal time abruptly transitions between different temporal valleys. 
Figs. \ref{fig6b} and \ref{fig6c} extract the 1D cross-sections along the horizontal and vertical axes. While the trajectory varies gently under weak coupling (Fig. \ref{fig6b}), breaching the critical threshold ($\approx 0.025\kappa$) instantaneously triggers violent oscillations and a precipitous drop. By taking optimal selection parameters $\Delta_{mc} = 7.0\kappa$ and $\gamma = 0.034\kappa$, the destructive interference is maximized, inducing a transient photon antibunching as deep as $10^{-9.43}$. This approach actively tailors the external environment rather than tuning the internal system.

Having discussed the $\lambda=0.01\kappa$ case in the previous figures, we now turn to the  $\lambda$-dependence of $g^{(2)}(0,t)$, as shown in Fig. \ref{fig7}. For the evolution up to $\kappa t=2$, $g^{(2)}(0,t)$ exhibits pronounced bunching, followed by antibunching over the subsequent time evolution. This arises from the delayed environmental feedback. At $\lambda=5.0\kappa$, $g^{(2)}(0,t)$ approaches the broad-spectrum limit, and is nearly identical to the result in the Markovian steady-state limit after $\kappa t=5$. A smaller $\lambda$ corresponds to a longer environmental memory time, which in turn gives more and deeper blockade dips. At $\lambda=0.01\kappa$, $g^{(2)}(0,t)$ drops to the order of $10^{-6}$. These results demonstrate that $\lambda$ provides a direct control over the memory timescale, and that prolonged environmental memory is essential for sustaining the transient non-Markovian blockade. 

So far, our discussion has been centered on the equal-time correlation $g^{(2)}(0,t)$. We now proceed to investigate the delayed second-order correlation function $g^{(2)}(t,t+\tau)$, which characterizes the temporal correlation between successively emitted single photons. In comparison with Fig. \ref{fig8b} the system exhibits distinctly different temporal dependencies and wave-packet profiles for $g^{(2)}(t_{opt},t_{opt}+\tau)$ in the transient non-Markovian state versus the Markovian state. Crucially, the delay time $\tau$ discussed here differs from the absolute evolution time $t$ in Fig. \ref{fig3a}. Here, $\tau$ represents the relative waiting time for the second photon after the first is detected. In the Markovian limit (Fig. \ref{fig8b}), the system reaches a steady state. Its correlation dynamics depend solely on the relative delay, yielding $g^{(2)}(\tau)$. It produces a narrow antibunching window, facilitating rapid blockade recovery suitable for high-frequency integrated quantum chips \cite{senellart2017}. Conversely, Fig. \ref{fig8a} reveals the significant advantage of the non-Markovian transient state in maintaining photon antibunching over an extended temporal window. Among the parameters presented in Fig. \ref{fig8a}, $\Omega_{m}=0.035\kappa$ is identified as the optimal choice. Correspondingly, when the first photon detection is locked at the optimal time $t=t_{opt}$ under this drive condition, the high-quality blockade window (the region below $g^{(2)}=0.5$) extends over a broad range of $\Delta(\kappa\tau)\approx 20$. This broad wave-packet feature can relax the timing jitter constraints in long-distance distributed quantum networks \cite{kimble2008}. However, this broadband blockade is highly sensitive to the driving conditions. As shown by the purple ($\Omega_{m}=0.010\kappa$) and green ($\Omega_{m}=0.200\kappa$) curves in Fig. \ref{fig8a}, deviations in drive strength destroy the destructive interference. This excites prominent bunching peaks or sidebands. Ultimately, by selecting the appropriate operation scheme (steady-state or transient) and precisely tuning the microwave parameters, we can flexibly tailor the temporal profile of emitted photons to meet the diverse interface requirements of quantum terminal networks.
\section{conclusion}
\label{sec:con}
In summary, we have theoretically proposed a hybrid optical-microwave crossed-cavity architecture based on an Er$^{3+}$:Y$_2$SiO$_5$ crystal in which a structured microwave reservoir actively engineers unconventional photon blockade through non-Markovian feedback. In this paper, we first compute the relevant characteristics of the structured reservoir - including the decoherence function, the time-dependent non-Markovian decay rate, and the renormalized microwave driving amplitude starting from the Volterra equotion. Within the time-convolutionless master-equation framework, we reveal that non-Markovian transient dynamics enable photon blockade performance significantly surpasses its steady-state counterpart. In addition, the microwave recycling terms cannot be neglected, thus the conventional wave-function amplitude approach is not applicable any more. In contrast to the traditional UPB \cite{li2019}, our structured microwave reservoir harnesses environmental information backflow to generate transient time-windows, which dynamically locks in the required destructive quantum interference. Furthermore, non-Markovian reservoirs offer a new degree of freedom for regulation: environmental parameters act as a new control knob, allowing us to actively manipulate the subject rather than being passively influenced. And such optimization does not necessarily require tuning microwave reservoirs, but via optical terminals. In experimental value, the delayed second-order correlation reveals the temporal constraints for single-photon synchronization and experimental robustness.

The present framework can be naturally extended to more general structured environments, including multimode reservoirs \cite{hc4r-2st8}, non-Lorentzian spectral densities \cite{Zhang:12}, and stronger coupling regimes beyond the perturbative limit \cite{PhysRevE.105.024126}. More broadly, the present work suggests that non-Markovianity can be engineered as an active control resource rather than merely a source of decoherence. Together with the rapid development of hybrid optical-microwave quantum interfaces \cite{bottcher2024,clerk2020}, they provides a feasible route toward experimentally realizing non-Markovian quantum interference and high-performance single-photon sources.
\begin{acknowledgments}
We thank Pengfei Yang for many stimulating discussions. This work was supported by the National Natural Science Foundation of China under Grants NO. 62575161 and NO. U21A20433.
\end{acknowledgments}
\section*{DATA AVAILABILITY}
The data that support the findings of this article are not publicly available. The data are available from the authors upon reasonable request.

\appendix
\section{ Derivation of the Non-Markovian Memory Kernel}

The non-Markovian feedback induced by the structured microwave reservoir is characterized by the symmetric environmental correlation function
\begin{equation}
	f(\tau)=\int_{0}^{\infty}d\Delta J(\Delta)\coth\left(\frac{\beta\hbar\omega_{mc}}{2}\right)\cos(\Delta\tau).
\end{equation}
Here the environment operators are expressed in the rotating frame $U_{B}=exp(i\omega_{m}t\sum_{k}b_{k}^{\dagger}b_{k})$, such that the environment correlation depends on the detuning from the microwave drive rather than the laboratory frequencies, which removes the fast oscillation. The thermal occupation factor $coth(\beta\hbar\omega_{mc}/2)$, however, remains dependent on the absolute environment frequency.

Here $J(\omega)=\sum_{k}|g_{k}|^{2}\delta(\omega-\omega_{k})$ is the spectral density, which parameterizes the coupling coefficients $g_{k}$. In this article, we focus on the following under-damped Brownian motion spectral density,
\begin{equation}
	J(\Delta)=\frac{2\gamma^{2}\Delta_{mc}\lambda\Delta}{(\Delta^{2}-\Delta_{mc}^{2})^{2}+\lambda^{2}\Delta^{2}}
\end{equation}
In the high-temperature regime relevant to the microwave reservoir, $\beta\hbar\omega\ll1$, the thermal factor can be approximated as
\begin{equation}
	\coth\left(\frac{\beta\hbar\omega_{mc}}{2}\right)\simeq\frac{2}{\beta\hbar\omega_{mc}}.
\end{equation}
Substituting this approximation into the correlation function and extending the integral to the whole frequency axis gives
\begin{equation}
	f(\tau)\simeq\frac{1}{2}\int_{-\infty}^{+\infty}d\Delta J(\Delta)\coth\left(\frac{\beta\hbar\omega_{mc}}{2}\right)e^{i\Delta\tau}.
\end{equation}
The integral can be evaluated by Cauchy's residue theorem. 

The dominant poles of the Lorentzian spectrum are located at
\begin{equation}
	\omega_{\pm}=\pm\Omega+i\lambda/2,
\end{equation}
where
\begin{equation}
	\Omega=\sqrt{\Delta_{mc}^{2}-\lambda^{2}/4}.
\end{equation}
Taking the residues at these poles yields the underdamped correlation function
\begin{equation}
	\begin{aligned}f(\tau) & =2\pi i\sum_{j=\pm}Res\left[\frac{2\gamma^{2}\omega_{mc}\lambda e^{i\Delta\tau}}{\beta\hbar\left[(\omega_{j}^{2}-\Delta{}_{mc}^{2})^{2}+\lambda^{2}\omega_{j}^{2}\right]},\omega_{j}\right]\\
		& =\frac{2\pi\gamma^{2}}{\beta\hbar\omega_{mc}}e^{-\frac{\lambda}{2}\tau}\left[\cos(\Omega\tau)+\frac{\lambda}{2\Omega}\sin(\Omega\tau)\right]
	\end{aligned}
\end{equation}
This oscillatory decaying kernel describes the finite-memory feedback of the structured microwave reservoir and reduces to the Markovian limit when the reservoir linewidth becomes sufficiently broad.
 \label{A}
 \section{  ODE and Perturbative Analytical Solution of the Decoherence Function $G(t)$}
 In non-Markovian dynamics, the system evolution is governed by a non-local Volterra integro-differential equation: $\dot{G}(t)=-\int_{0}^{t}f(t-\tau)G(\tau)d\tau.$ Direct solution is difficult due to this historical convolution term. In this section, we first recast this equation via Laplace transform into a set of closed ordinary differential equations (ODEs). This serves as a  benchmark for numerical integration. Subsequently, we apply a first-order perturbative approximation under the weak-coupling condition to derive its analytical expression.
 \subsection*{B1. Laplace Transform of the Memory Kernel}
 We apply the Laplace transform $\tilde{f}(s)=\int_{0}^{\infty}f(\tau)e^{-s\tau}d\tau$ to the correlation function $f(\tau)$ derived in Eq. \eqref{8}. Utilizing the standard formulas $\mathcal{L}\{e^{-at}\cos(bt)\}=\frac{s+a}{(s+a)^{2}+b^{2}}$ and $\mathcal{L}\{e^{-at}\sin(bt)\}=\frac{b}{(s+a)^{2}+b^{2}}$, we simplify the expression to obtain:
  \begin{equation}
 	\begin{aligned}
 \tilde{f}(s)&=K\left[\frac{s+\lambda/2}{(s+\lambda/2)^{2}+\Omega^{2}}+\frac{\lambda}{2\Omega}\frac{\Omega}{(s+\lambda/2)^{2}+\Omega^{2}}\right]
 \\
 &=K\frac{s+\lambda}{(s+\lambda/2)^{2}+\Omega^{2}}
 	\end{aligned}
 \tag{B1}
\end{equation}
 Similarly, applying the Laplace transform to the Volterra equation yields $s\tilde{G}(s)-1=-\tilde{f}(s)\tilde{G}(s)$. Rearranging this directly produces Eq. \eqref{10}
  \subsection*{B2.  ODE Solution}
  We define an auxiliary variable $\tilde{y}(s)$:
  \begin{equation}
  \tilde{y}(s)\equiv\tilde{f}(s)\tilde{G}(s)=\frac{K(s+\lambda)}{s^{2}+\lambda s+\omega_{c}^{2}}\tilde{G}(s)
  \tag{B2}
  \label{B2}
  \end{equation}
  Expanding Eq. \eqref{B2} and eliminating the denominator, we perform the inverse Laplace transform ($s\rightarrow d/dt$) back to the time domain. This yields a second-order differential equation:
  \begin{equation}
  \ddot{y}(t)+\lambda\dot{y}(t)+\omega_{c}^{2}y(t)=K\dot{G}(t)+K\lambda G(t). 
  \tag{B3}
  \label{B3}
  \end{equation}
  From the initial definition, $s\tilde{G}(s)-1=-\tilde{y}(s)$ maps to $\dot{G}(t)=-y(t)$. Substituting this into Eq. \eqref{B3} recovers Eq. \eqref{11} .To reduce the differential order, we introduce the state vector $\mathbf{X}(t)=[G(t),y(t),z(t)]^{T}$, where $z(t)=\dot{y}(t)$. Eq. \eqref{B3} can then be equivalently written as:
  \begin{equation}
 \frac{d}{dt}\begin{bmatrix}G\\y\\z\end{bmatrix}=\begin{bmatrix}0 & -1 & 0\\0 & 0 & 1\\K\lambda & -(\omega_{c}^{2}+K) & -\lambda\end{bmatrix}\begin{bmatrix}G\\y\\z\end{bmatrix}. 
 \tag{B4}
 \label{B4}
 \end{equation}
 Given the initial condition $\mathbf{X}(0)=[1,0,K]^{T}$, Eq. \eqref{B4} can be numerically integrated using standard algorithms, providing the reference for the numerical solutions presented in Fig. \ref{fig1}.
 \subsection*{B3. First-Order Perturbative Solution}
 Expanding Eq. \eqref{10}, we obtain:
  \begin{equation}
 \tilde{G}(s)=\frac{1}{s+\tilde{f}(s)}=\frac{s^{2}+\lambda s+\omega_{c}^{2}}{s(s^{2}+\lambda s+\omega_{c}^{2})+K(s+\lambda)}. 
 \tag{B5}
\end{equation}
 Let $N(s)=s^{2}+\lambda s+\omega_{c}^{2}$ denote the numerator polynomial. The decoherence dynamics are dictated by the characteristic equation from the denominator, $P(s)=sN(s)+K(s+\lambda)=0$. The solution permits a partial fraction expansion: $\tilde{G}(s)=\sum_{j}\frac{C_{j}}{s-s_{j}}$, where $s_{j}$ are the poles of $P(s)$. The inverse transform naturally yields $G(t)=\sum_{j}C_{j}e^{s_{j}t}$. Here, $C_{j}$ is the residue corresponding to pole $s_{j}$, evaluated as:
 \begin{equation}
 C_{j}=\operatorname{Res}[\tilde{G}(s),s_{j}]=\frac{s_{j}^{2}+\lambda s_{j}+\omega_{c}^{2}}{\prod_{k\neq j}(s_{j}-s_{k})}. 
 \tag{B6}
 \label{B6}
 \end{equation}
 Under the weak-coupling condition $K\ll\omega_{c}^{2}$, we expand both poles and residues into first-order power series in $K$: $s_{j}=s_{j}^{(0)}+Ks_{j}^{(1)}+\mathcal{O}(K^{2})$ and $C_{j}=C_{j}^{(0)}+KC_{j}^{(1)}+\mathcal{O}(K^{2})$. Substituting these expansions into $G(t)$ gives: $G(t)=\sum_{j}\bigl(C_{j}^{(0)}+KC_{j}^{(1)}\bigr)e^{s_{j}^{(0)}t}e^{Ks_{j}^{(1)}t}+\mathcal{O}(K^{2}).$Applying the Taylor expansion $e^{Ks_{j}^{(1)}t}=1+Ks_{j}^{(1)}t+\mathcal{O}(K^{2})$, we ultimately obtain:
  \begin{equation}
  	\footnotesize
 G(t)=\sum_{j}C_{j}^{(0)}e^{s_{j}^{(0)}t}+K\sum_{j}\Bigl(C_{j}^{(1)}e^{s_{j}^{(0)}t}+C_{j}^{(0)}s_{j}^{(1)}te^{s_{j}^{(0)}t}\Bigr)+\mathcal{O}(K^{2}). 
 \tag{B7}
\end{equation}
 Next, we evaluate the zeroth-order terms and first-order corrections. By treating $K(s+\lambda)$ as a first-order perturbation to the bare polynomial $P_{0}(s)=s(s^{2}+\lambda s+\omega_{c}^{2})$, we identify the unperturbed poles: the Markovian principal pole $s_{0}^{(0)}=0$ and the conjugate sideband poles $s_{\pm}^{(0)}=-\frac{\lambda}{2}\pm i\Omega$. Substituting $s_{j}\approx s_{j}^{(0)}+Ks_{j}^{(1)}$ into $P(s_{j})=0$ yields $P_{0}(s_{j}^{(0)})+K\left[s_{j}^{(1)}P_{0}'(s_{j}^{(0)})+P_{1}(s_{j}^{(0)})\right]+\mathcal{O}(K^{2})=0$. Since the zeroth-order poles inherently satisfy $P_{0}(s_{j}^{(0)})=0$, we extract the first-order pole correction:
  \begin{equation}
 s_{j}^{(1)}=-\frac{P_{1}(s_{j}^{(0)})}{P_{0}'(s_{j}^{(0)})}=-\frac{s_{j}^{(0)}+\lambda}{3(s_{j}^{(0)})^{2}+2\lambda s_{j}^{(0)}+\omega_{c}^{2}}. 
 \tag{B8}
\end{equation}
From Eq. \eqref{B6}, the zeroth-order residues are readily evaluated as $C_{0}^{(0)}=1$ and $C_{\pm}^{(0)}=0$. By applying a similar first-order perturbative expansion to Eq. \eqref{B6}, the residue corrections $C_{j}^{(1)}$ can be derived. Specifically, for the sideband poles, this procedure yields:
\begin{equation}
	C_{\pm}^{(1)}=\frac{s_{\pm}^{(1)}N'(s_{\pm}^{(0)})}{\bigl(s_{\pm}^{(0)}-s_{0}^{(0)}\bigr)\bigl(s_{\pm}^{(0)}-s_{\mp}^{(0)}\bigr)}. 
	\tag{B9}
\end{equation}

Notably, the initial condition $G(0)=1$ constrains the overall normalization. This dictates that the first-order amplitude correction of the principal pole satisfy $C_{0}^{(1)}=-2\operatorname{Re}[C_{+}^{(1)}]$, which corresponds to a initial amplitude slip. Because it represents a mere constant perturbation and does not alter the long-time exponential decay, we safely neglect it. Combining the perturbative solutions and order matching above, the analytical decoherence function is reconstructed as: $G(t)=e^{-\frac{K\lambda}{\omega_{c}^{2}}t}+K\left[\frac{2\Omega-i\lambda}{\Omega(2\Omega+i\lambda)^{2}}e^{(-\frac{\lambda}{2}+i\Omega)t}+\frac{2\Omega+i\lambda}{\Omega(2\Omega-i\lambda)^{2}}e^{(-\frac{\lambda}{2}-i\Omega)t}\right]+\mathcal{O}(K^{2})$, obtained in Eq. \eqref{12}.
\label{B}
	\begin{figure}[htbp] 
	\centering
	\includegraphics[width=0.42\textwidth]{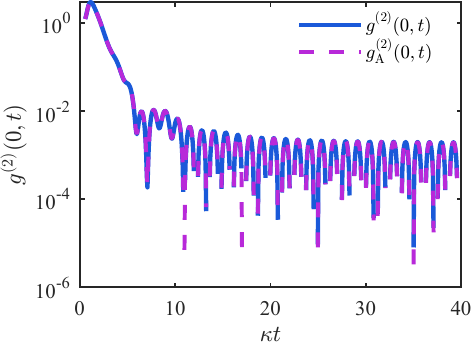}
	\caption{ Comparison between the full master-equation result $g_{{\rm ME}}^{(2)}(0,t)$ (solid blue line) and the fourth-order perturbation $g_{{\rm rec}}^{(2)}(0,t)$ (dashed purple line). The close agreement over the full evolution. }
	\label{fig9}
\end{figure}
\begin{figure}[htbp] 
	\hspace{-1.5cm}
	\centering
	\subfloat[]{
		\includegraphics[width=0.26\textwidth]{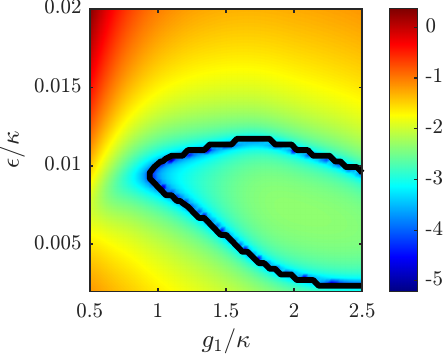}
		\label{fig10a}
	}
	\subfloat[]{
		\includegraphics[width=0.26\textwidth]{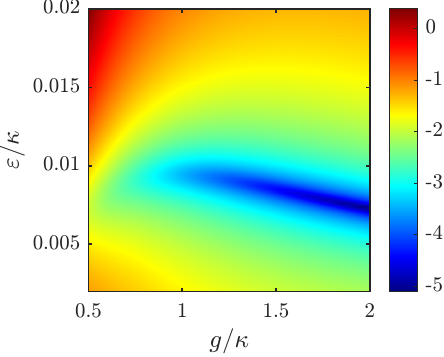}
		\label{fig10b}
	}
	\caption{$g^{(2)}(0,t)$ in the space of atom-cavity coupling $g/\kappa$ and probe amplitude $\epsilon/\kappa$. (a) Results using the full master equation including the microwave recycling terms. The black line represents the analytical solution. (b) The corresponding results obtained after removing the microwave recycling terms.}
	\label{fig10}
\end{figure}	
\section{ Density-Matrix Formulation and the Role of Microwave Recycling}
\label{D}
In the weak-probe expansion, the density matrix is decomposed as $\rho(t,\epsilon)=\sum_{j}\epsilon^{j}\rho^{(j)}(t),\rho^{(j)}(t)=\frac{1}{n!}\frac{\partial^{n}\rho(t,\varepsilon)}{\partial\varepsilon^{n}}_{\varepsilon=0}$. Substituting this expansion into the master equation, the zeroth-order term satisfies $\rho^{(0)}(t)=\mathcal{U}_{0}(t,0)\rho(0)$, while higher orders follow the recursive relation $\rho^{(j)}(t)=\int_{0}^{t}d\tau\mathcal{U}_{0}(t,\tau)\mathcal{V}\rho^{(j-1)}(\tau)$. $\mathcal{U}_{0}(t,\tau)$ is the propagator generated by $L_{0}(t)$. Using this hierarchy up to fourth order yields the approximate result $g_{{\rm A}}^{(2)}(0,t)$ shown in Fig. \ref{fig9}.
	
Figure \ref{fig10a} presents the blockade map of ${g^{(2)}(0,t)}$ in the (g,$\varepsilon$) plane obtained from the full master equation, including the microwave recycling terms. The black boundary is obtained from the fourth-order perturbative expansion.  For comparison, Fig. \ref{fig10b} presents the corresponding dynamics after removing these recycling terms; the transient bifurcated structure disappears entirely, demonstrating that the microwave recycling terms are indispensable for reproducing the transient blockade and therefore the density-matrix description cannot be omitted.

	% --- 参考文献 ---
	\section*{References}
	\nocite{*}
	\bibliography{reference} 
	
\end{document}